\documentstyle[version2,aps]{revtex} 
\begin{document}                                                        
\draft                                                                  
\twocolumn
\widetext
\begin{title}
Spin and charge dynamics of the 2D t-J model at intermediate
electron densities:\\ 
absence of spin-charge separation 
\end{title}

\author{R. Eder and Y. Ohta}
\begin{instit}
Department of Applied Physics, Nagoya University, Nagoya 464-01, Japan
\end{instit}
\begin{abstract}
We present an exact diagonalization study of the dynamical spin and 
density correlation functions in small clusters of $t-J$ model, focussing 
on the regime of intermediate and low electron densities, $\rho_e$$<$$0.5$.
In 2D both correlation functions agree remarkably well with the convolution 
of the single-particle spectral function, i.e. the simplest estimate possible
within a Fermi liquid picture. Deviations from the convolution are shown to 
originate from symmetry-related selection rules, which are unaccounted for 
in the convolution estimate. For all fillings under consideration, we show 
that the low energy peaks originate from particle-hole excitations between 
the Fermi momenta, as expected for a Fermi liquid. We contrast this with the 
behaviour in $1D$, where spin and density correlation function show the 
differences characteristic of spin-charge separation and where neither 
correlation function is approximated well by the convolution.
\end{abstract} 
                                                                               
\narrowtext                  
\topskip9.25cm
A much discussed issue in the theory of high-temperature superconductivity 
is the question
whether strongly correlated electron models in $2D$ share the most 
striking feature of their $1D$ counterparts, spin charge separation. 
Whereas in a Fermi liquid the low lying spin and density excitations 
can be modelled by particle-hole 
(i.e. composite) excitations of the fermionic quasiparticles, which represent 
the `most elementary' excitations of the system, the situation is reversed 
in a spin-charge separated system. There, the soliton-like spin and density 
excitations are themselves the elementary excitations, and the physical 
electrons can be considered as composite excitations.\\
By numerical diagonalization one can obtain the exact excitation spectra
of small clusters and should thus, at least in principle, be able
to decide whether they are consistent with any given theoretical
scenario. We therefore have performed a study of the 
dynamical spin and density correlation functions of the
$t$$-$$J$ model, which reads
\[
  H =                                                    
 -t \sum_{< i,j >, \sigma}                                         
( \hat{c}_{i, \sigma}^\dagger \hat{c}_{j, \sigma}  +  H.c. )
 + J \sum_{< i,j >}[\;\bbox{S}_i \cdot
 \bbox{S}_j
 - \frac{n_i n_j}{4}\;].
\]
The $\bbox{S}_i$ are the electronic spin operators, 
$\hat{c}^\dagger_{i,\sigma} =c^\dagger_{i,\sigma}(1-n_{i,-\sigma})$
and the sum over $<i,j>$ stands for a summation       
over all pairs of nearest neighbors. All results to be presented
below have been obtained for $1D$ and $2D$ $16$ site clusters,
with the parameter value $J/t$$=$$0.4$. In the present study we restrict 
ourselves to intermediate and low electron densities, where the results
are more easily understood than near half-filling\cite{Tohyama}. 
To be more precise, in this doping regime
the excitation spectrum of the $2D$ $t$$-$$J$ model 
turns out to be completely consistent with the Fermi-liquid derived
particle-hole picture, even deviations from the noninteracting case
are small. Our results suggest that throughout the range of dopings
studied, the $t$$-$$J$ model represents a not even very strongly
correlated Fermi liquid. No detail of our results would necessitate the 
introduction of the concept of spin charge separation.\\ 
Using the standard Lanczos algorithm we computed the dynamical spin (SCF) 
and density (DCF) correlation functions:
\[
C_\alpha(\bbox{q},\omega) 
= \frac{1}{\pi} \Im \langle \Psi_0 | O_\alpha^\dagger
\frac{1}{ \omega - ( H- E_0 ) - i0^+} O_\alpha | \Psi_0\rangle.
\]  
Here $|\Psi_0\rangle$ ($E_0$) denotes the ground state
wave function (ground state energy), and
as the operator $O$ we choose the Fourier transform of either the
density operator $n_{i,\uparrow} + n_{i,\downarrow}$ 
($\alpha=d$) or the spin operator
$(1-\rho_e) (n_{i,\uparrow} - n_{i,\downarrow})$ ($\alpha=s$). 
We introduced an extra prefactor of $2(1-\rho_e)$ for the
$z$-spin operator because with this definition both correlation 
functions obey the same sum-rule,
\[
\sum_{\bbox{q}\neq 0} \int_0^{\infty} d\omega C_\alpha(\bbox{q},\omega)
= \rho_e (1 - \rho_e ),
\] 
which faciliates their comparison. 
In a system of noninteracting particles, the above two particle
Green's functions can be expressed as the convolution of the single 
particle photoemission (PES) and inverse photemission (IPES) spectrum,
i.e.
\[
C_c(\bbox{q},\omega) = \frac{2}{N^2} \sum_{\bbox{k}} \int  d\omega'
A_{PES}(\bbox{k},\omega') A_{IPES}(\bbox{q}-\bbox{k},\omega-\omega').
\]
Having computed the PES and IPES spectra by the Lanczos method
we can replace the $\delta$-peaks in these functions by Lorentzians and 
obtain the convolution numerically.
Thereby the brodening of the $\delta$-peaks in the spectral function
was taken $1/2$ of the one for the correlation 
functions, because the numerical convolution effectively 
\newpage
\topskip0cm
\noindent
doubles the width of the peaks. The convolution obeys the sum rule
\[
\sum_{\bbox{q}\neq 0} \int_0^{\infty} d\omega C_c(\bbox{q},\omega)
= \rho_e (1 - \rho_e ) - \frac{2}{N^2} \sum_{\bbox{k}}
n_{\bbox{k}} ( m(\delta) - n_{\bbox{k}}),
\]
where $n(\bbox{k})= \langle \hat{c}_{\bbox{k},\sigma}^\dagger
\hat{c}_{\bbox{k},\sigma}\rangle$ denotes the ground state
momentum distribution and $m(\delta)= (1+\delta)/2$ with $\delta$ the
hole concentration. The last term is different from $0$ but turns out to 
be quite small, so that all spectral functions under consideration have 
a comparable integrated weight.\\
Approximating the spin and density correlation function
by a mere convolution is a drastic approximation, which neglects
for example all vertex corrections; as we will 
show now, for the electron densities under consideration
it is nevertheless a remarkably good approximation.
Figs. \ref{fig1}-\ref{fig4} compare the convolution estimate
with the exact SCF and DCF for electron densities between
$10/16$ and $4/16$. While the agreement for $6$ holes is not yet
very impressive (particularly for the SCF),
the agreement becomes better and better with decreasing
electron density. With the exception of the low energy peak structure at
$(\pi,0)$ and $(\pi/2,\pi/2)$ (which is determined by symmetry-related
selection rules unaccounted for in the convolution, see below) 
the overall shape of the exact correlation
functions is reproduced quite well, sometimes to the degree of a
one-to-one correspondence between the dominant features.
While SCF and DCF are not obviously identical, they both
can be viewed as being derived by slight but different
modifications of the convolution. \\
We now turn to the low energy peaks at $(\pi,0)$ and
$(\pi/2,\pi/2)$, indicated by arrows in Figs. \ref{fig2}-\ref{fig4}. 
Obviously these are not reproduced well by the convolution.
We will show that this is the consequence of point group
selection rules, which are not accounted for in the convolution.
Assuming the existence of a free electron like Fermi surface,
the $6$ hole ground state would correspond to a closed-shell configuration,
where the momenta $(0,0)$, $(\pm\pi/2,0)$ and $(0,\pm pi/2)$ are
occupied. The ground states with lower electron densities
then would have `holes' in the outer shell of momenta, which are therefore
only partially occupied. This is confirmed by the spectral
function at $(\pi/2,0)$, which shows only a PES peak immediately below 
the Fermi energy for $10$ electrons, but both a PES peak immediately below 
and an IPES peak immediately above the Fermi energy for $8$, $6$ and $4$ 
electrons. The low energy peaks in the convolution
for momentum transfer $(\pi,0)$ and $(\pi/2,\pi/2)$ then originate
precisely from transitions within this shell of partially occupied
momenta $(\pm\pi/2,0)$, $(0,\pm\pi/2)$, as indicated in Fig. \ref{fig4}a. 
We adopt the hypothesis that the lowest states
of the system can be described by a `cluster version' of
Fermi liquid theory. To be more precise, we assume that e.g. for the
$8$ electron case there is a one-to-one mapping of e.g. the lowest states 
of the cluster and `quasiparticle states'
of the type $a_{\bbox{k},\uparrow} a_{\bbox{k}',\downarrow}
|FS\rangle$, where  $a_{\bbox{k},\sigma}$ annihilates a quasiparticle,
$|FS\rangle$ denotes the closed shell Fermi sea of
$10$ noninteracting electrons and $\bbox{k}$ and $\bbox{k}'$
are restricted to the shell of momenta $(\pm\pi/2,0)$, $(0,\pm\pi/2)$.
We next assume that there is a residual interaction between the 
quasiparticles, which lifts the degeneracy between the various 
states of this type with given total momentum. 
To simplify the notation we label the momenta as
indicated in Fig. \ref{fig4}b. Then,
from the fact that the exact $8$ electron
ground state is a spin singlet and has $d_{x^2-y^2}$ symmetry, we 
conclude that it should be modelled by the state
\begin{equation}
\frac{1}{2} ( 
a_{1,\uparrow} a_{3,\downarrow}
-a_{1,\downarrow} a_{3,\uparrow}
- a_{2,\uparrow} a_{4,\downarrow}
+ a_{2,\downarrow} a_{4,\uparrow} ) |FS\rangle.
\end{equation}
This would also be consistent with the pronounced
$d_{x^2-y^2}$ pairing correlations found for larger values of
$J/t$ at this electron density\cite{DagottoRiera,Ohtaetal}.
Next, the only states in our subspace with momentum $(\pi/2,\pi/2)$ are
\begin{equation}
\frac{1}{\sqrt{2}} ( 
a_{3,\uparrow} a_{4,\downarrow}\pm
a_{3,\downarrow} a_{4,\uparrow}) |FS \rangle.
\label{state1}
\end{equation}
The antisymmetric combination is a spin singlet and
even under reflection by the $(1,1)$ direction, so that its 
point group symmetry is incompatible with that of the
ground state: there can be no peak in the DCF.
The symmetric combination is a spin triplet and has an acceptable 
point group symmetry
so that we expect a peak in the SCF. Both predictions are consistent 
with the exact spectra.  
The only possible state with momentum $(\pi,0)$ and the required point
group symmetry (even under reflections by both $x$ and $y$ axis)
in the model space is
\begin{equation}
\frac{1}{\sqrt{2}} (a_{1,\uparrow} a_{1,\downarrow} 
+ a_{3,\uparrow} a_{3,\downarrow}) |FS\rangle.
\end{equation}
This is a spin singlet so that the SCF cannot have a low energy peak. 
Since the point-group symmetry of this state is consistent with the
selection rules, we expect a low-energy peak in the DCF.
Again both predictions are consistent with the exact spectra, although
the weight of the low energy peak in the DCF is extremely small.\\
Next, literally the same explanation holds for the case of $4$ electrons,
if we replace the annihilation operators by creation operators,
choose the fully occupied $\Gamma$-point as the `Fermi sea'
and replace $(3,4)$$\rightarrow$$(1,2)$ in (\ref{state1}). 
This is because
like the $8$ electron ground state, the $4$ electron ground state is a 
spin singlet with $d_{x^2-y^2}$ symmetry. We thus expect a low 
energy peak in the DCF at $(\pi,0)$ and in the SCF at $(\pi/2,\pi/2)$; 
inspection of the exact spectra shows, that this is indeed realized.\\   
We turn to the case of $6$ electrons. Here we have to use quasiparticle 
states with $4$ holes in the $10$-electron closed shell. The exact
ground state has unusual quantum numbers: its total spin is $S=2$ 
and it belongs to the $B_2$ (or $d_{xy}$) representation of the $C_{4v}$ 
point group. This suggests to model the $S_z=-2$ member of the ground 
state multiplet by the state
$a_{1,\uparrow}a_{2,\uparrow}a_{3,\uparrow}a_{4,\uparrow}|FS\rangle$, which
obviously has the correct point group symmetry. The $S_z=0$ member of the
ground state multiplet (which we are considering) can now in principle
be obtained by acting twice with the spin raising operator.
It is easy to see, however, that all necessary conclusions can already 
be drawn from the final states: introducig the operator
$t_{i,j} = 1/\sqrt{2} ( a_{i,\uparrow} a_{j,\downarrow} +
a_{i,\downarrow} a_{j,\uparrow} )$, which creates a spin triplet
on the momenta $i$ and $j$, the only state with
total momentum $(\pi,0)$ and the required point group symmetry
(odd under reflection by both $x$- and $y$-axis) reads
\begin{equation}
\frac{1}{\sqrt{2}} ( a_{1,\uparrow} a_{1,\downarrow} -
a_{3,\uparrow} a_{3,\downarrow}) t_{2,4} |FS\rangle.
\end{equation}
This is a spin triplet, so that we expect a peak in the SCF
but not in the DCF. 
Similarly, the state with maximal spin and momentum $(\pi/2,\pi/2)$
that is even under reflection by the $(1,1)$ direction reads
\begin{equation}
\frac{1}{\sqrt{2}} (a_{3,\uparrow} a_{3,\downarrow} t_{1,4}  +
a_{4,\uparrow} a_{4,\downarrow} t_{2,3}) |FS\rangle.
\end{equation}
Again, this is a triplet, so that we expect a peak in the SCF but not
in the DCF. All predictions are consistent with the numerical
spectra.\\ 
We thus have shown that the low energy peak structure can be understood 
completely
by a simple `cluster Fermi liquid theory', which relies on nothing more 
than elementary selection rules. To further strengthen the evidence for 
this interpretation, we now compare the electron momentum distribution
$n(\bbox{k}) = \langle c_{\bbox{k},\sigma}^\dagger
c_{\bbox{k},\sigma} \rangle$ 
in the ground state (GS) to that of the final states (FS) associated 
with the low-energy peaks in the SCF and DCF, i.e. we consider 
\begin{equation}
\Delta n(\bbox{k}) = n_{FS}(\bbox{k}) - n_{GS}(\bbox{k}).
\label{deldef}
\end{equation}
Based on the above `model wave functions', we can predict the changes of 
the quasiparticle occupation numbers which accompany the respective 
spin or density excitation.
These are listed in Table \ref{tab1}. Up to a constant factor
due to the wave function renormalization constant $Z$, the
exact $\Delta n(\bbox{k})$ then should be the same.\\
To obtain the final state wave functions,
we take advantage of the fact that the energies of the low energy
peaks give highly precise estimates for the eigenenergies $E_{final}$
of the SCF or DCF final states. Thus, by applying powers of
$(H - E_{final})^{-1}$ to some randomly chosen trial state
(thereby employing the conjugate gradient algorithm)
we can efficiently converge out the respective final state
wave function\cite{InverseIteration} and obtain its momentum 
distribution.
Then,  Figs. \ref{fig7}-\ref{fig9} show the $\Delta n(\bbox{k})$
for all low lying peaks in both SCF and DCF.
By comparison with Table \ref{tab1} it can be seen that the exact
results are indeed completely consistent with the particle-hole picture:
the changes are always substantially larger for the Fermi momenta than
for any other momentum, 
the reduction of the magnitude of $\Delta n(\bbox{k})$ as compared 
to the `free electron' estimate suggests values of the quasiparticle 
weight $Z$ between $0.6$ (for $8$ electrons) and $0.8$ 
(for $4$ electrons). The losses of $n(\bbox{k})$ at the Fermi momenta 
always exceed the gains; this is probably due to an enhanced scattering 
in the final (=excited) states. We thus have rather
unambiguous evidence for the validity of the Fermi liquid like 
particle-hole picture at low frequencies. For higher frequency the 
good agreement between the exact correlation functions and the 
convolution strongly suggests the same.\\ 
We now turn to a comparison with the $1D$ model, where
spin charge separation rather than Fermi liquid
behaviour is known to be realized at all doping 
levels\cite{BaresBlatter}.
Fig. \ref{fig5} compares the SCF and DCF with the
convolution of the single particle spectral function for
$\rho_e$$=$$0.5$. The Fermi momentum is
$k_F$$=$$\pi/4$, and SCF and DCF show their lowest energy excitation
at different momenta,
$2k_F$$=$$\pi/2$ and $4k_F$$=$$\pi$, respectively, which is 
characteristic of spin charge separation\cite{BaresBlatter}. 
Unlike in $2D$ neither correlation function is approximated well 
by the convolution. The latter 
consists almost entirely of structureless continua, whithout the 
sharp peaks of the exact correlation functions.\\
For the low density regime our results are consistent with those 
of Jagla {\em et al.}\cite{Jagla}. These authors studied the group 
velocity of spin and charge excitations in small clusters and found 
different velocities for spin and charge in $1D$ (indicating spin 
charge separation) but identical velocities (indicating absence of 
spin charge separation) in $2D$.\\
On the other hand,
evidence for spin charge separation at electron densities 
$0.75$ and $0.2$ was seen by 
Putikka {\em et al.}\cite{Putikka} in the static SCF and DCF
of the $2D$ $t$$-$$J$ model obtained by high-temperature expansion.
However, a subsequent cluster study of the same quantities by 
Chen {\em et al.}\cite{Chen} could not strengthen this 
evidence: the DCF showed no indication of a singularity
at the `characteristic wavevector' for spinless Fermions.  
Moreover the {\em dynamical} SCF and DCF
differ drastically for low and high doping\cite{Tohyama} so that
it seems not very plausible that they both can be described by the
same physics. Our results show that while for low electron density
the peak structure of SCF and DCF is similar and 
consistent with the convolution, there are differences in the pole 
strengths. Thus, while SCF and DCF apparently can be explained by 
particle-hole transitions between analogous states, the matrix elements
are different. This would support an explanation of the
the static correlation functions e.g. by RPA 
calculations for the large-$U$ Hubbard Model\cite{Bulut}.\\
In summary, we have studied the spin and density corelation function
for intermediate and low electron densities in the $t$$-$$J$ model.
For a spin-charge separated system, (i.e. in $1D$),
these two correlation functions cannot adequately be described in 
a particle-hole excitation picture. Contrary to this, our results for 
$2D$ show that the simplest estimate based on the particle-hole 
excitation picture, namely the convolution of the single particle 
spectral function, provides a remarkably good description of these 
correlation 
functions. The low energy peak structure can be completely
explained by a simple `cluster Fermi liquid theory' which relies only
on the particle-hole picture and elementary selection rules.
The respective particle-hole transitions can be directly made visible in
the momentum distribution. Our exact results thus suggest that for the 
less than quarter filled case the $t$$-$$J$ model has a fairly 
conventional particle-hole type excitation spectrum, and represents a 
not even very strongly correlated Fermi liquid. 
There is no need to invoke spin-charge separation to understand the 
dynamics of the $2D$ $t$$-$$J$ model for the electron 
densities under consideration.\\
It is interesting to note that for electron densities near half-filling 
(i.e. $2$ and $4$ holes in the $4\times4$ cluster) the situation 
changes completely: the DCF consists almost entirely of incoherent,
high-energy continua, the SCF has sharp low energy peaks and almost 
no continua\cite{Tohyama}. The dominant spin excitation at
$(\pi,\pi)$ is a spin-wave like collective mode, i.e. a remnant of the 
undoped system; only spin excitations with different momentum transfer
still correspond to particle-hole transitions\cite{spins}.
In fact, quite a number of physical quantities show a significant change
at hole concentration $\sim$$0.2$-$0.3$: the doping dependence 
of the Drude weight\cite{Dagottohub}, 
the temperature dependence of the susceptibility\cite{TohyamaI},
the sign of the Hall constant\cite{CastilloBalseiro}.
For $J$$=$$0$ Chiappe {\em et al.}\cite{Chiappe} 
have shown that the overlap of the exact cluster ground state
with Gutzwiller wave functions drops sharply from $\sim$$1$ for low 
and intermediate electron densities to $\sim$$0$ for high density.
The clear distinction which can be seen in a wide variety of
physical quantities suggests to assume two quite different phases
for the $2D$ $t$$-$$J$ model as a function of doping: 
for $\rho_e<0.7$$-$$0.8$ there seems to be a fairly 
conventional Fermi liquid, with a ground state 
that is `adiabatically connected' to the noninteracting one and 
correlations are of little importance. 
Near half-filling, the correlations dominate and the system seems to be
continuous with the insulator: the spectral function shows rigid-band 
behaviour\cite{rigid} with increasing doping, the Fermi surface takes 
the form of hole pockets\cite{Moreo2,pockets}, the spin excitation
spectrum is reminiscent of the undoped 
Heisenberg antiferromagnet\cite{spins}.\\ 
It is a pleasure to acknowledge numerous instructive discussions
with Professor S. Maekawa and Dr. T. Tohyama.
Financial support of R.E. by the Japan Society for the Promotion
of Science is most gratefully acknowledged. Parts of the
computations were carried out at the Institute for Molecular
Science, Okazaki.
\figure{Spin correlation function (right column, dotted line)
and density correlation function (left column, dotted line)
as compared to the convolution of the single particle spectral
function (full line). All spectra are for
the $4\times 4$ cluster $t$$-$$J$ model with $10$ electrons and
$J/t=0.4$.
\label{fig1}}
\figure{Same as Fig. \ref{fig1} for $8$ electrons.
\label{fig2}}
\figure{Same as Fig. \ref{fig1} for $6$ electrons.
\label{fig3}}
\figure{Same as Fig. \ref{fig1} for $4$ electrons.
\label{fig4}}
\figure{Possible particle-hole transitions between the `Fermi momenta' 
in the $4\times4$ cluster (a), labelling of the `Fermi momenta' (b).
\label{fig5}}
\figure{Difference of $n(\bbox{k})$ for
the final states belonging to the low energy
peaks indicated by arrows in Fig. \ref{fig2} and $n(\bbox{k})$
fort the ground state i.e. the quantity
$\Delta n(\bbox{k})$ defined in (\ref{deldef}). 
The figure shows all momenta
in the Brillouin zone, the `Fermi momenta' of Fig. \ref{fig5} 
are indicated by dark symbols, the top right corner corresponds to
$(\pi,\pi)$. The number of electrons is $8$.
\label{fig6} }
\figure{Same as Fig. \ref{fig6} for $6$ electrons.
\label{fig7}}
\figure{Same as Fig. \ref{fig6} for $4$ electrons.
\label{fig8}}
\figure{a: Comparison of convolution (full line) and exact 
DCF (dotted line) for the $1D$ $16$-site chain with $8$ electrons
and $J/t=0.4$.\\
b: same as (a) but for the SCF.
\label{fig9}}
\begin{table}
\caption{Change of the electron occupation numbers of the
Fermi momenta as obtained from the model wave functions
for the SCF and DCF final states.}
\begin{tabular}{c c c | c c c c}
$N_e$ & $\bbox{k}$ & $\alpha$ 
& $(\frac{\pi}{2},0)$ & $(0,\frac{\pi}{2})$ & 
$(-\frac{\pi}{2},0)$ & $(0,-\frac{\pi}{2})$ \\
\hline
8 & $(\pi,0)$&  DCF &$ -0.25$ &$ +0.25$ &$ -0.25$ &$ +0.25$ \\
8 & $(\frac{\pi}{2},\frac{\pi}{2})$ & SCF &$ +0.25$ &$ +0.25$ &$ -0.25$ 
&$ -0.25$ \\
6 & $(\pi,0)$ & SCF &$\;\;\;0.00$ &$\;\;\;0.00$ &$ \;\;\;0.00$ &$ 
\;\;\;0.00$ \\
6 & $(\frac{\pi}{2},\frac{\pi}{2})$ & SCF &$ +0.25$ &$ +0.25$ &$ -0.25$ 
&$ -0.25$ \\
4 & $(\pi,0)$ & DCF &$ +0.25$ &$ -0.25$ &$ +0.25$ &$ -0.25$ \\
4 & $(\frac{\pi}{2},\frac{\pi}{2})$ & SCF &$ +0.25$ &$ +0.25$ &$ -0.25$
 &$ -0.25$ 
\end{tabular}
\label{tab1}
\end{table}
                                                        
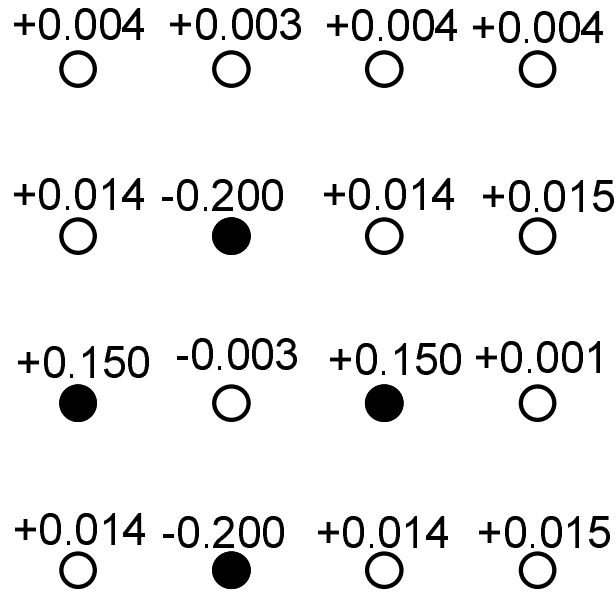
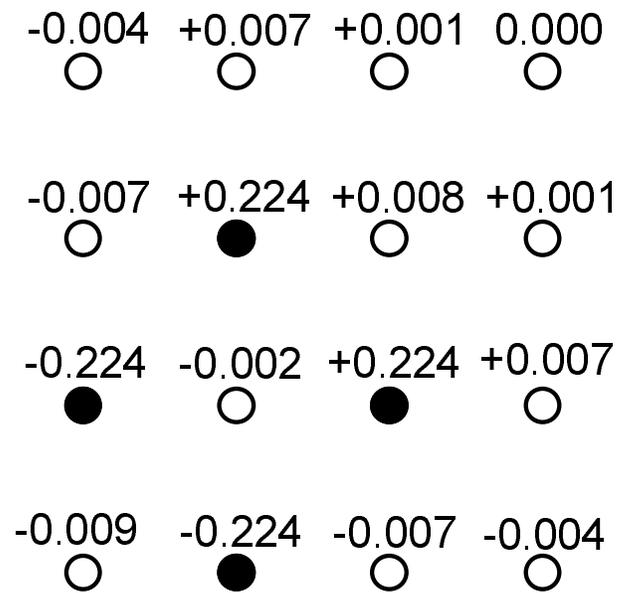
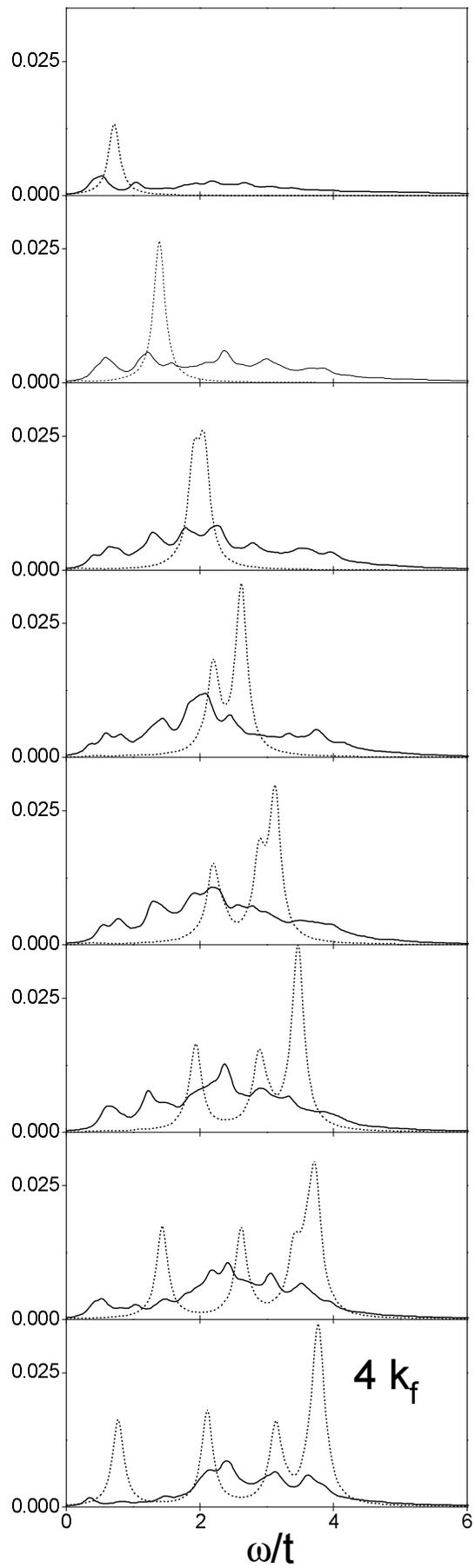
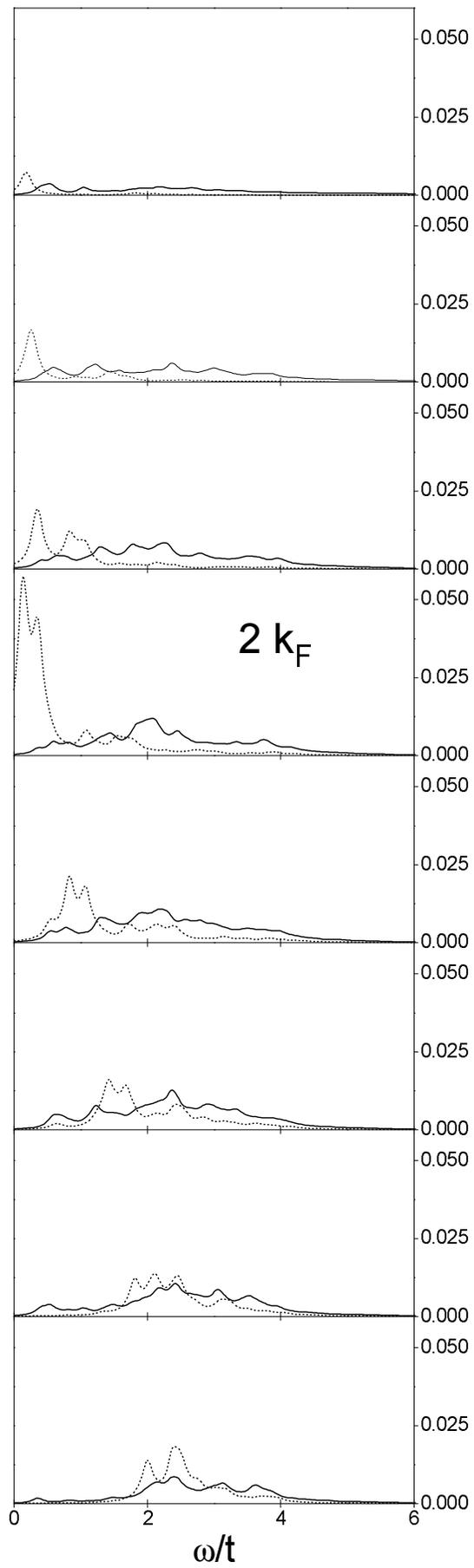
\end{document}